\pdfoutput=1
\documentclass[aps,prb,reprint,amsmath,amssymb]{revtex4-1}

\usepackage{graphicx}
\usepackage{dcolumn}
\usepackage{bm}
\usepackage[utf8]{inputenc}
\usepackage{subcaption}
\usepackage{optidef}
\usepackage{float}
\usepackage{xcolor}

\def\b#1{{\mathbf{#1}}}

\begin{document}



\title{Inferring crystal electronic properties from experimental data sets through Semidefinite Programming}


\author{Benjamin De Bruyne}
 \email{Corresponding author: \mbox{benjamin.debruyne@centraliens.net}}
 \affiliation{%
CentraleSupélec School, Paris-Saclay University, 91190 Gif-sur-Yvette, France
}%
 \altaffiliation[Also at ]{School of Engineering, Liège University, Belgium}
 \author{Jean-Michel Gillet}%
 \affiliation{%
   Structures, Properties and Modeling of Solids Laboratory, CentraleSupélec School, CNRS-UMR8580, Paris-Saclay University, 91190 Gif-sur-Yvette, France
}%

\date{\today}

\begin{abstract}
 Constructing a quantum description of crystals from scattering experiments is of paramount importance to explain their macroscopic properties and to evaluate the pertinence of theoretical \textit{ab-initio} models. While reconstruction methods of the one-electron reduced density matrix have already been proposed, they are usually tied to strong assumptions that limit and may introduce bias in the model. The goal of this paper is to infer a one-electron reduced density matrix (1-RDM) with minimal assumptions. We have found that the mathematical framework of Semidefinite Programming can achieve this goal. Additionally, it conveniently addresses the nontrivial constraints on the 1-RDM which were major hindrances for the existing models. The framework established in this work can be used as a reference to interpret experimental results. This method has been applied to the crystal of dry ice and provides very satisfactory results when compared with periodic \textit{ab-initio} calculations. 
\end{abstract}


\maketitle

\section{Introduction}
The computation of one-electron expectation values such as the mean position, the mean momentum or the mean kinetic energy of electrons in a crystal does not require more than the mere knowledge of the one-electron reduced density matrix (1-RDM) \cite{lowdin55,coleman63,mcweeny60,davidson76}. This quantity provides a quantum description of an average electron and has been proved to be sufficient \cite{lathiotakis08,gilbert75}. Furthermore, the electron density in position and momentum spaces can easily be derived from such a quantity. It is therefore a useful tool for describing electronic properties at a quantum level. Additionally, using the 1-RDM is well suited to represent mixed states systems using statistical ensembles of pure states. This is generally the case for crystals at non-zero temperature.

Several models have been proposed to approximate and refine a 1-RDM from experimental expectation values \cite{deutsch12,deutsch14,hansen78,gillet01,gillet07,gillet04,gillet99,gueddida18,pillet01,schmider92,clinton72,tsirelson96}. The complementarity between position and momentum space expectation values in the description of the 1-RDM is now well accepted \cite{cooper04,pisani12}. For this reason, deep inelastic X-ray scattering data known as ``directional Compton scattering profiles'' (DCPs), have been taken into account in addition to  X-ray or polarized neutron diffraction structure factors (SFs) to refine a variety of models. The former are related to 2D projections of electron density in momentum space, while the latter are linked to the Fourier coefficients of the electron density in position space. However, almost all of these models require an initial guess or assumption on the electronic configuration. When these are inappropriate or too simple, there is a risk that the model, hence the results, will be affected by a severe bias. 
The purpose of this work is to investigate and assess a new method to obtain a 1-RDM from expectation values with minimal bias.

In order to serve as a reference, an initial periodic \textit{ab initio} calculation (at the DFT level) has been conducted from which the reference 1-RDM was extracted. From the same calculation, a limited number of structure factors and directional Compton profiles were generated. Once a random noise was added, these deteriorated data constituted our pseudo-experimental data.

The method explicitly takes into account the so called \textit{N-representability} conditions \cite{coleman63}, which ensure that the inferred 1-RDM is quantum mechanically acceptable, i.e. that there exists a many-electron wavefunction from which the 1-RDM can be derived. Addressing these nontrivial conditions is made possible by the use of Semidefinite Programming \cite{boyd96}, a recent subfield of convex optimization \cite{boyd04} which is of growing interest in Systems \& Control Theory, Geometry and Statistics \cite{wolkowicz12}.

 
\section{Method}
\label{sec:method}

\subsection{Molecular spin-traced 1-RDM}
In the following section, for simplicity, we will restrict our treatment to a crystal with a single molecule per cell that has $N$ paired electrons. The method can be generalized to several molecules by either assigning a 1-RDM to each molecule provided that they can be considered electronically isolated from each other (as in Sec.\ref{sec:results}), or defining one 1-RDM for a group of interacting molecules. Additionally, spin-orbitals can be employed to construct two spin resolved 1-RDMs when the system bears unpaired electrons.

Let  $\{\chi_i\}_{i\in\{1,\ldots,n\}}$ be a set of atomic orbitals describing the electrons of each atom taken as an independent system. From $\{\chi_i\}_{i\in\{1,\ldots,n\}}$, one can deduce an orthogonal basis set $\{\phi_i\}_{i\in\{1,\ldots,n\}}$ for the molecule, using Löwdin orthogonalization procedure \cite{lowdin50} for example.  Expanding the spin-traced 1-RDM $\widehat{\Gamma} (\b{r},\b{r}')$ in such a basis, one reveals its basis set representation: the population matrix $\mathrm{\widehat{P}}$, so that: 
\begin{align}
  \widehat{\Gamma} (\b{r},\b{r}') = \sum_{i,j}^{n} \mathrm{\widehat{P}_{ij}} \phi^*_i(\b{r}) \phi_j(\b{r'})
  \label{eq:pop}
\end{align}
Although it is not necessary to use an orthogonal basis, it is done here because the \textit{N-representability} conditions are conveniently expressed in such a basis. In general, these conditions are expressed on the eigenvalues of the spin-traced 1-RDM. In this case, they are translated into conditions on the eigenvalues of $\mathrm{\widehat{P}}$ and state that they must lie in $[0,2]$ (as $N$ is even) and their sum must be equal to $N$. 

\subsection{Expectation values}
\label{sec:expectVal}

Any one-electron expectation value $\langle \widehat{O} \rangle$ can be calculated from its operator $\mathcal{ \widehat{O}_{\b{r}'}}$ applied to the 1-RDM $\widehat{\Gamma}(\b{r},\b{r'})$: 
\begin{align}
  \langle \widehat{O} \rangle = \int \left(\mathcal{ \widehat{O}}_{\b{r}'}\ \widehat{\Gamma}(\b{r},\b{r'})\right)_{\b{r'}=\b{r}}  d\b{r} 
\end{align}
where $\mathcal{ \widehat{O}_{\b{r}'}}$ means that the operator only acts on variable $\b{r}'$.
By defining, the basis set representation of $\mathcal{ \widehat{O}}$ as:
\begin{align}
  \mathrm{ \widehat{O}_{ij}} = \int\phi^*_i(\b{r})\ \left(\mathcal{ \widehat{O}}_{\b{r'}}\ \phi_j(\b{r'})\right)_{\b{r'}=\b{r}}  d\b{r}
\end{align}
one can conveniently write the expectation value as $\langle  \widehat{O} \rangle=\mathrm{tr(\widehat{P}\widehat{O})}$ (using Eq.\ref{eq:pop}), where $\mathrm{tr}$ is the matrix trace operator.

In particular, in position space, the X-ray structure factors $ F(\b{q})$, which are given by: 
\begin{align}
  F(\b{q}) =& \int \widehat{\Gamma}(\b{r},\b{r}) e^{i\b{q}\cdot\b{r}} d\b{r} \label{eq:F}\\
           =&\sum_{i,j}^n \mathrm{\widehat{P}_{ij}}  \int  \phi_i^*(\b{r}) \phi_j(\b{r}) e^{i\b{q}\cdot\b{r}} d\b{r} 
\end{align}
have an operator whose basis set representation is:
\begin{align}
  \mathrm{\widehat{F}_{ij}(\b{q})} =& \int \phi^*_i(\b{r}) \phi_j(\b{r}) e^{i\b{q}\cdot\b{r}} d\b{r} 
\end{align}

In momentum space, the directional Compton profiles $J^{\b{u}}(q)$ can be defined through the autocorrelation function $B(\b{r})$ \cite{pattison79,weyrich79,benesch71} as:
\begin{align}
  J^{\b{u}}(q) =& \int \frac{1}{2\pi}  B(t\b{u})  e^{-i t\, q} dt \label{eq:J}\\
  B(\b{r}) =& \int \widehat{\Gamma}(\b{r'},\b{r+r'})  d\b{r'} \label{eq:B}
\end{align}
Their operator basis set representation is therefore: 
\begin{align}
  \mathrm{\widehat{J}_{ij}^{\b{u}}(q)} = \int \int  \frac{1}{2\pi}   \phi^*_i(\b{r'}) \phi_j(t\b{u}+\b{r'}) e^{-i t\, q}\, dt\, d\b{r'}
\end{align}
From Eq.\ref{eq:F} and Eq.\ref{eq:J}-\ref{eq:B}, one can appreciate the complementarity of both expectation values as they, respectively, shed light upon the diagonal and the off-diagonal directions of the 1-RDM.

\subsection{Constrained least-squares fitting scheme}

In the Bayesian sense, the objective is to infer the most probable population matrix $\mathrm{\widehat{P}}$  so that it fits given independent expectation values $\langle \widehat{O}_\alpha \rangle$. In the following, the expectation values $\langle \widehat{O}_\alpha \rangle$ are SFs and DCPs data. Supposing the latter follow Gaussian error distributions with standard deviations $\sigma_\alpha$ and no \textit{a priori} knowledge is given on $\mathrm{\widehat{P}}$, the problem is equivalent to minimizing the so-called $\chi^2$ function with respect to the elements of $\mathrm{\widehat{P}}$ \cite{gillet04,sivia06}. It can be summarized in the following optimization program: 

\begin{mini}{\mathrm{\widehat{P}}}{ \sum_\alpha \frac{1}{\sigma_\alpha^2}\left| \langle \widehat{O}_\alpha \rangle - \mathrm{tr(\widehat{P}\widehat{O}_\alpha)}\right|^2}
{\label{eq:S}}{}
\addConstraint{\mathrm{tr(\widehat{P})}}{= N}
\addConstraint{\mathrm{\widehat{P}}}{ \succcurlyeq 0}
\addConstraint{\mathrm{2\, I-\widehat{P}}}{ \succcurlyeq 0}
\end{mini}
where  $\mathrm{I}$ is the identity matrix  and the notation $\mathrm{A} \succcurlyeq  0$ means that $\mathrm{A}$ is a symmetric positive semi-definite matrix, i.e.\ its eigenvalues are non-negative. The last two constraints are mathematically equivalent to the condition that the eigenvalues of $\mathrm{\widehat{P}}$ must lie in $[0,2]$.

The following passage will cast program (\ref{eq:S}) as a semidefinite optimization program. These steps are quite standard in the field of convex optimization \cite{boyd04}. Introducing a new variable $t$, program (\ref{eq:S}) is equivalent to:
 \begin{mini}
   {\mathrm{\widehat{P}},t}{t}
   {\label{eq:s}}{}
   \addConstraint{\mathrm{tr(\widehat{P})}}{= N}
   \addConstraint{\mathrm{\widehat{P}}}{ \succcurlyeq 0}
   \addConstraint{\mathrm{2\, I-\widehat{P}}}{ \succcurlyeq 0}
   \addConstraint{t - ||\b{\Delta_\sigma O}||^2}{ \geq 0}
\end{mini}
where $\b{\Delta_\sigma O}$ is a column vector whose elements are  $(\langle \widehat{O}_\alpha \rangle   - \mathrm{tr(\widehat{P}\widehat{O}_\alpha)})/ \sigma_\alpha$ and $||\cdot||$ is the euclidean norm.

Using Schur's complement \cite{zhang06}, the last constraint of program (\ref{eq:s}) can be written as a \textit{linear} \textit{matrix inequality}: 
\begin{align}
  \left[\begin{array}{cc}
          \mathrm{I} &  \b{\Delta_\sigma O} \\
          (\b{\Delta_\sigma O})^T & t
\end{array}\right] & \succcurlyeq  0 
\end{align}
where $\mathrm{I}$ is the identity matrix of appropriate dimensions. This inequality is indeed linear with respect to $\widehat{\mathrm{P}}$ as $\b{\Delta_\sigma O}$ is a linear function of $\widehat{\mathrm{P}}$.

This type of program where the objective function is linear and the constraints are linear combinations of symmetric matrices that must be positive semidefinite, has been extensively studied and is referred to as the class of Semidefinite Programming \cite{boyd96}. Interior-point algorithms can be used to solve this class of problems and no initial guess is required. Treatment of the 2-RDM by Semidefinite Programming has already been reported in the context of variational computation of molecules \cite{mazziotti07}.

In the present work, this program has been addressed by using the optimization software \textsc{Mosek} \cite{mosek} interfaced by \textsc{Yalmip} toolbox \cite{Lofberg2004} under \textsc{Matlab}.%

\section{Application to dry ice}
\label{sec:results}

\begin{figure}[h]
\includegraphics[width=0.35\textwidth]{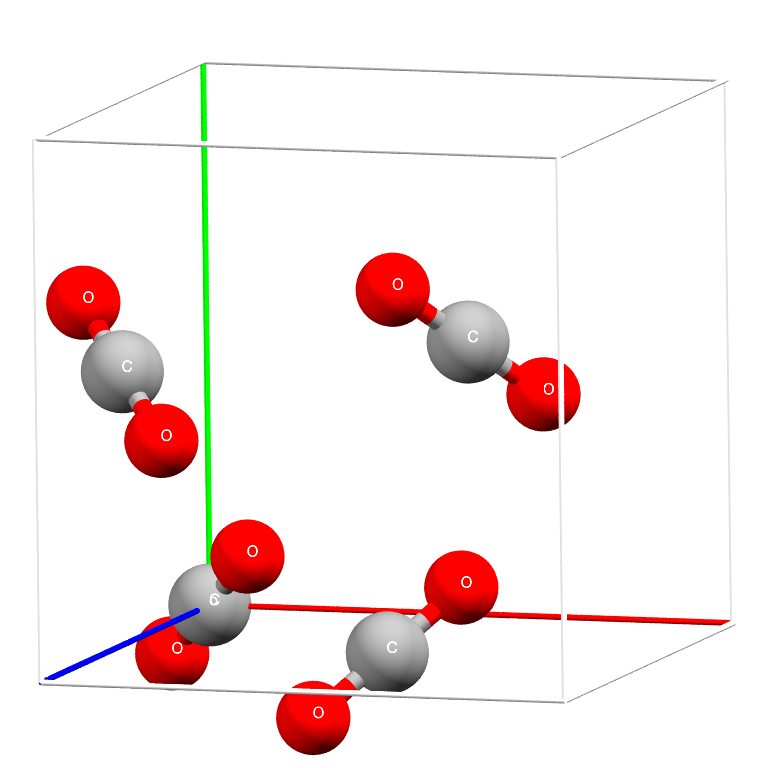}
\caption{Unit cell of dry ice: space group $Pa\bar{3}$, $a=5.63$~\AA \cite{deSmedt1924}. }
\label{fig:CO2}
\end{figure}

Dry ice $\mathrm{CO_2}$ is a molecular crystal with four molecules per cubic unit cell (Fig.\ref{fig:CO2}).

\subsection{Expectation values generation}
For the following example, structure factors and directional Compton profiles have been generated using the \textsc{Crystal14} periodic \textit{ab-initio} software \cite{DovesiOrlando14,DovesiSaunders14}. Density Functional Theory and the B3LYP of hybrid exchange and correlation functional have been chosen as a theoretical framework. Large polarized and diffuse atomic basis sets (triple-zeta valence with polarization quality) \cite{peintinger13,civalleri12} for both types of atoms have been used.

In the following, $1800$ structure factors ($(h,k,l)_{\text{cubic cell}} \in \mathrm{Z}^3\, |\,  0\leq h \leq 7,  -7\leq k \leq 7,  -7\leq l \leq 7$, $\sin(\theta_{max})/\lambda\sim 1.08$ \AA$^{-1}$) and three directional Compton profiles ($u=(h,k,l)_{\text{cubic cell}} \in \{(0,0,1),(1,1,0), (1,1,1)\}$), with a resolution of $0.15$ a.u. and limited to $6$ a.u. were computed.

\begin{figure}[h]
\includegraphics[width=0.5\textwidth]{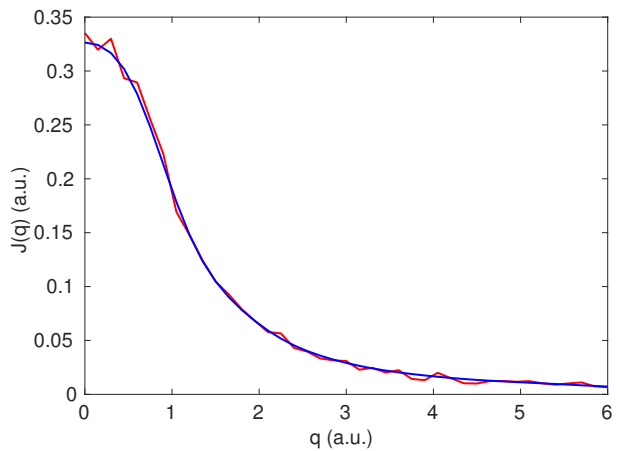}
\caption{Directional Compton profile $J^{\b{u}}(q)$ (in red) and $\mathrm{tr(\widehat{P}\widehat{J}^{\b{u}}(q))}$ (in blue) for dry ice in the crystallographic direction $\b{u}=(1,1,1)$ in the conventional cell. The spectrum is in atomic units and the profile is normalized to one electron.}
\label{fig:CPDiff} 
\end{figure}

\begin{figure}
    \centering
    \begin{subfigure}[t]{0.26\textwidth}
        \includegraphics[width=\textwidth]{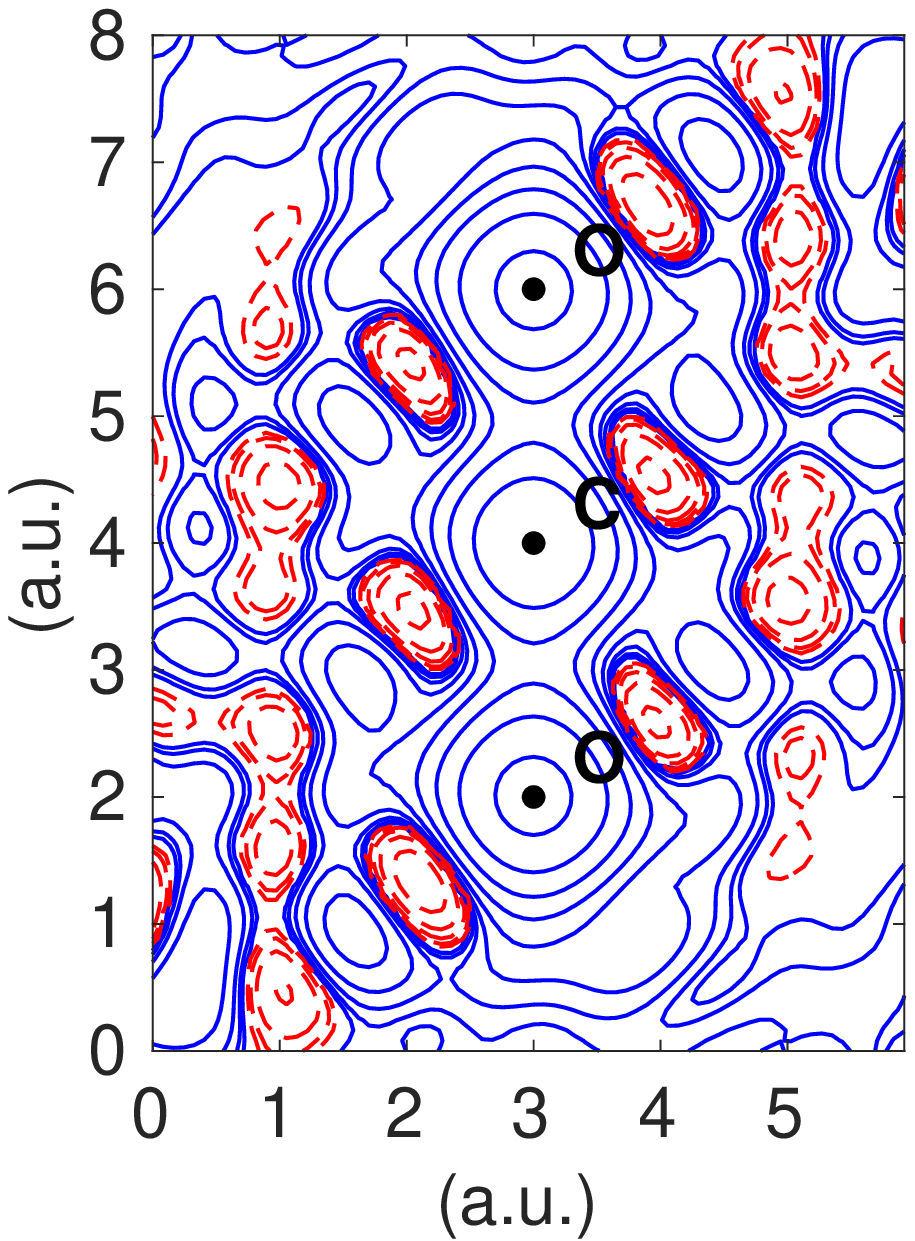}
    \end{subfigure}\hspace{-10mm}
    \begin{subfigure}[t]{0.26\textwidth}
        \includegraphics[width=\textwidth]{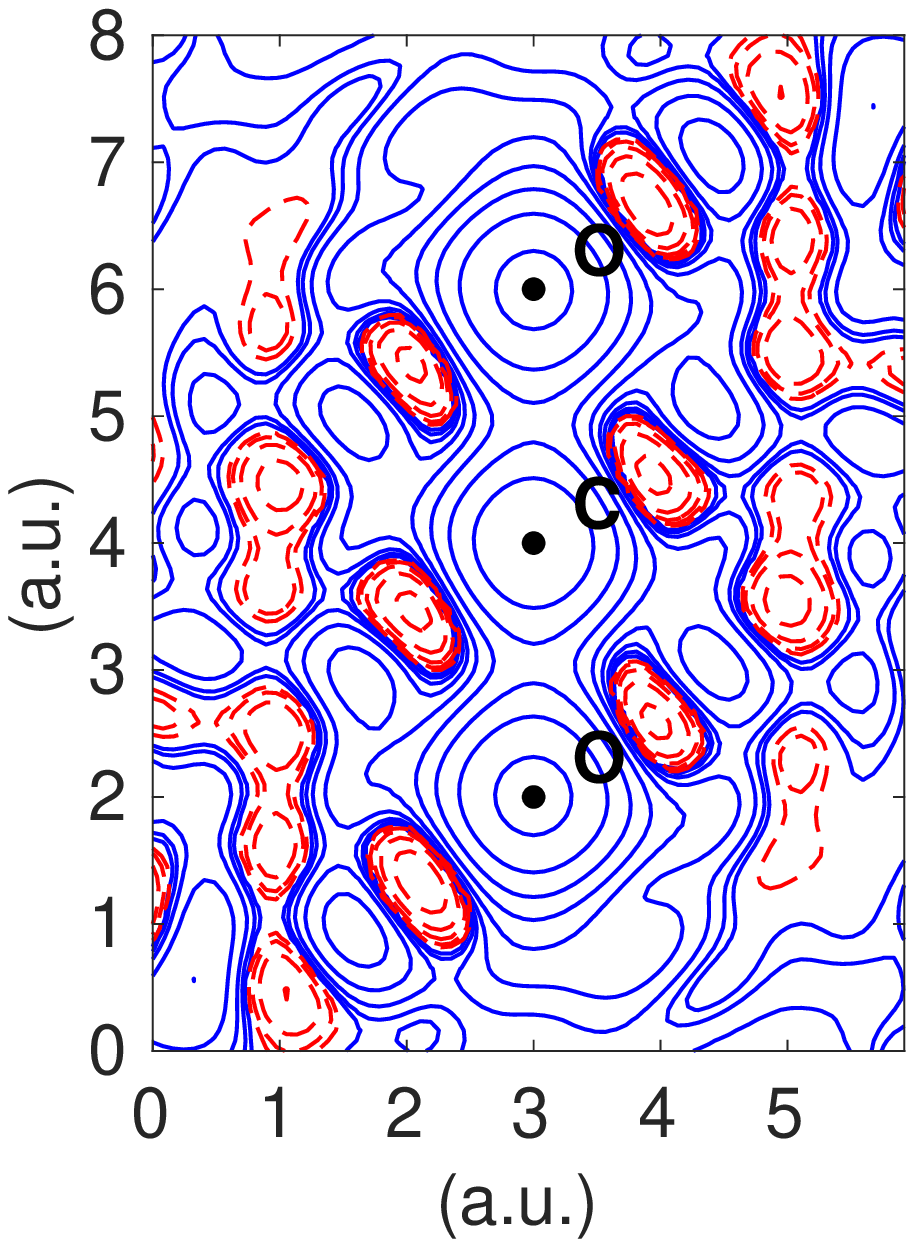}
      \end{subfigure}
    \caption{Density map reconstructed from truncated Fourier series with coefficients $\mathrm{tr(\widehat{P}\hat{F}(\b{q}))}$ (left) and $F(\b{q})$ (right) in a plane including the O-C-O bonding. Contours at intervals of $\pm 0.01\times 2^n$ a.u.$^{-3}$ ($n=0$-$20$): positive and negative contours \cite{footnoteFourier} are blue solid lines and red dashed lines respectively.}\label{fig:Fourier}
  \end{figure}

To prove the robustness of the method, Gaussian errors have been added to the data. For each structure factor, the standard deviation is $3\%$ of its modulus and for each directional Compton profile $J^{\b{u}}(q)$, it is set to be $\sqrt{ J^{\b{u}}(q)/\alpha_{\b{u}}}$ where $\alpha_{\b{u}}$ is such that $\sqrt{ J^{\b{u}}(0)/\alpha_{\b{u}}}=0.03 \times J^{\b{u}}(0)$. Such distorted DCPs and SFs are illustrated respectively in Fig.\ref{fig:CPDiff} and  in Fig.\ref{fig:Fourier} by means of a Fourier density map. In the following, the resulting distorted DCPs and SFs will be qualified as ``pseudo-experimental'' data.

\begin{figure*}
    \centering
    \begin{subfigure}[t]{0.30\textwidth}
        \includegraphics[width=\textwidth]{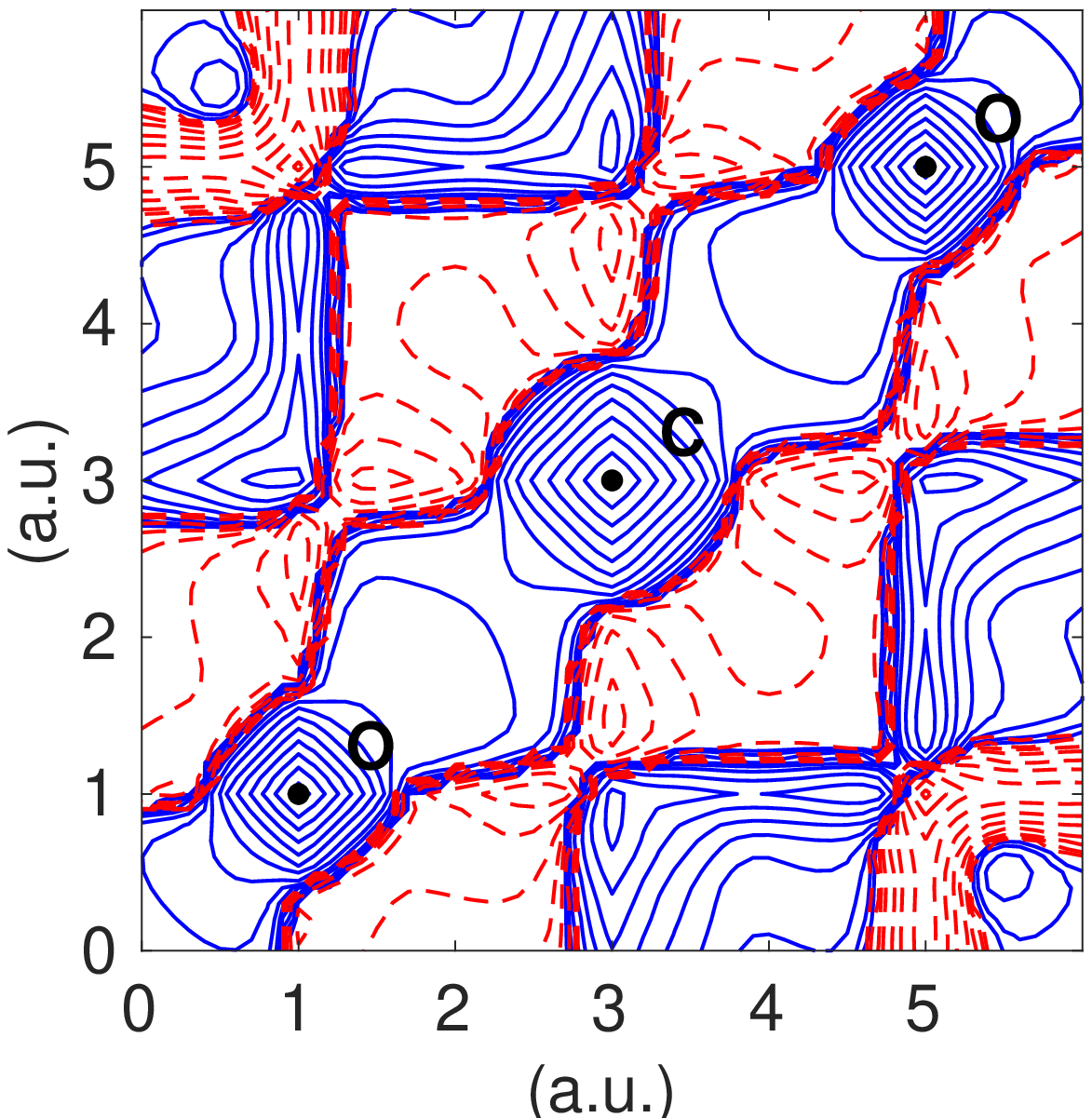}
        \caption{}
        \label{fig:RDMSDP}
    \end{subfigure}
    \begin{subfigure}[t]{0.30\textwidth}
        \includegraphics[width=\textwidth]{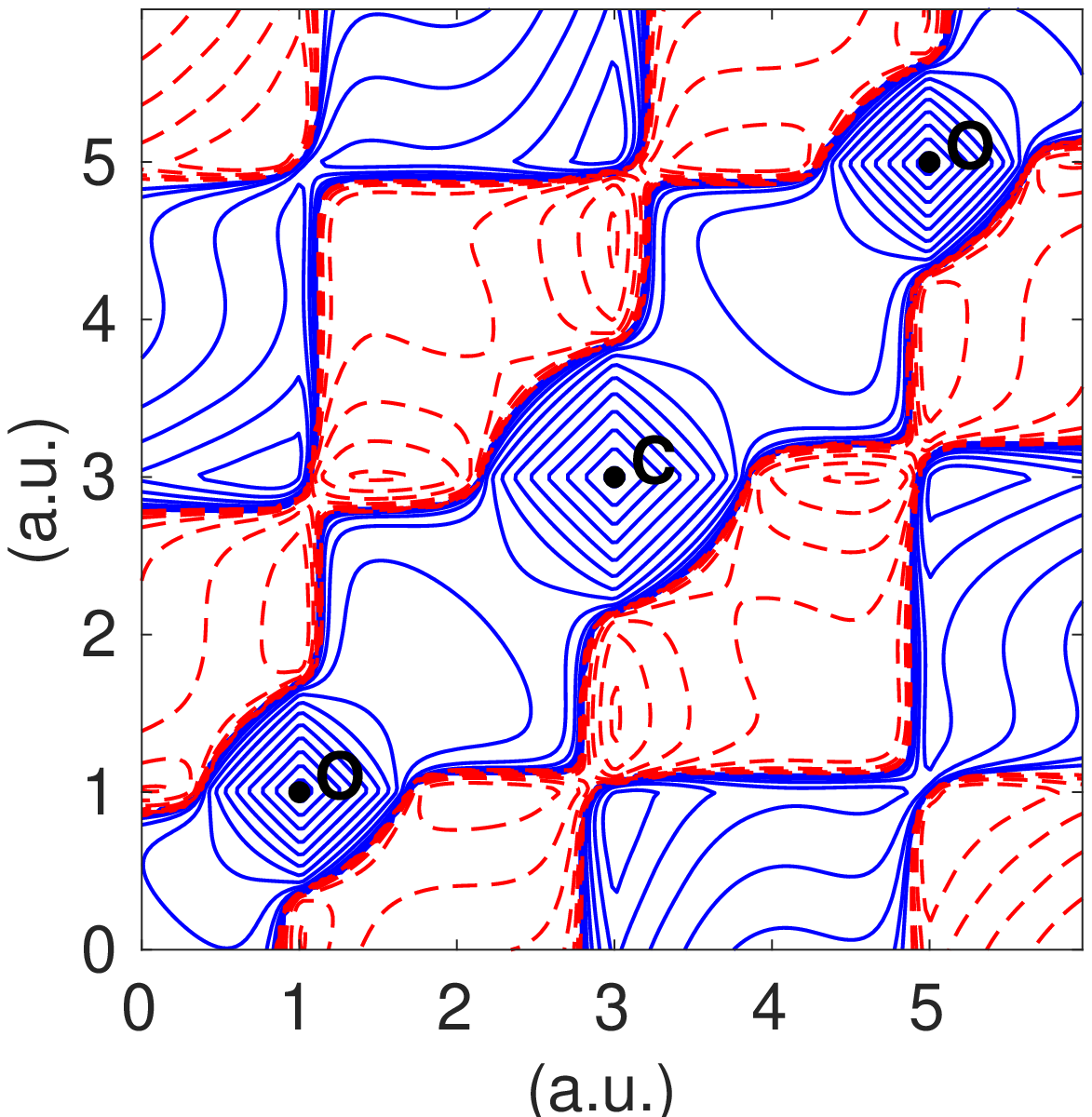}
        \caption{}
        \label{fig:RDMCrystal}
      \end{subfigure} \\
       \begin{subfigure}[t]{0.30\textwidth}
        \includegraphics[width=\textwidth]{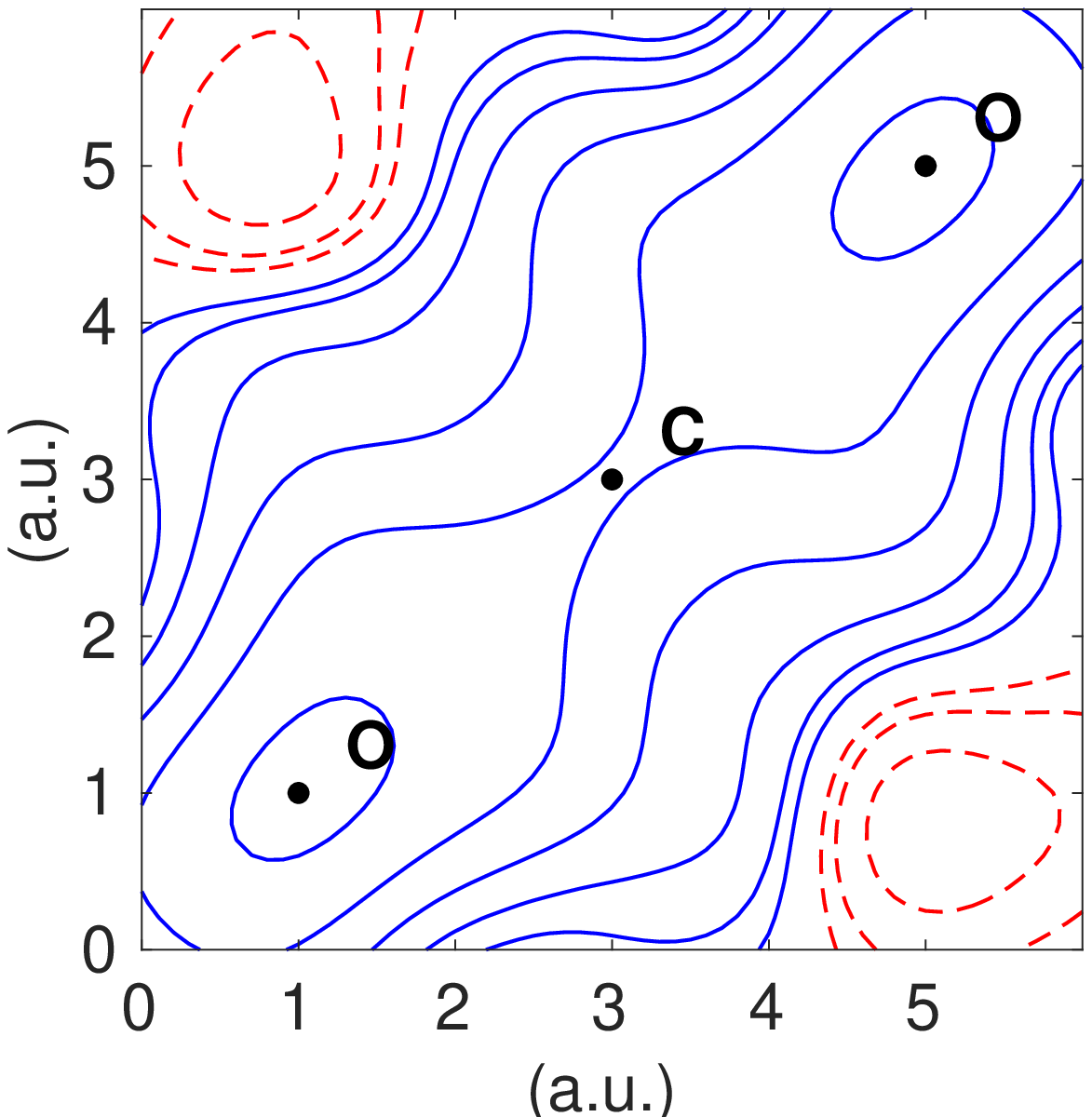}
        \caption{}
        \label{fig:RDMSDPUp}
    \end{subfigure}
    \begin{subfigure}[t]{0.30\textwidth}
        \includegraphics[width=\textwidth]{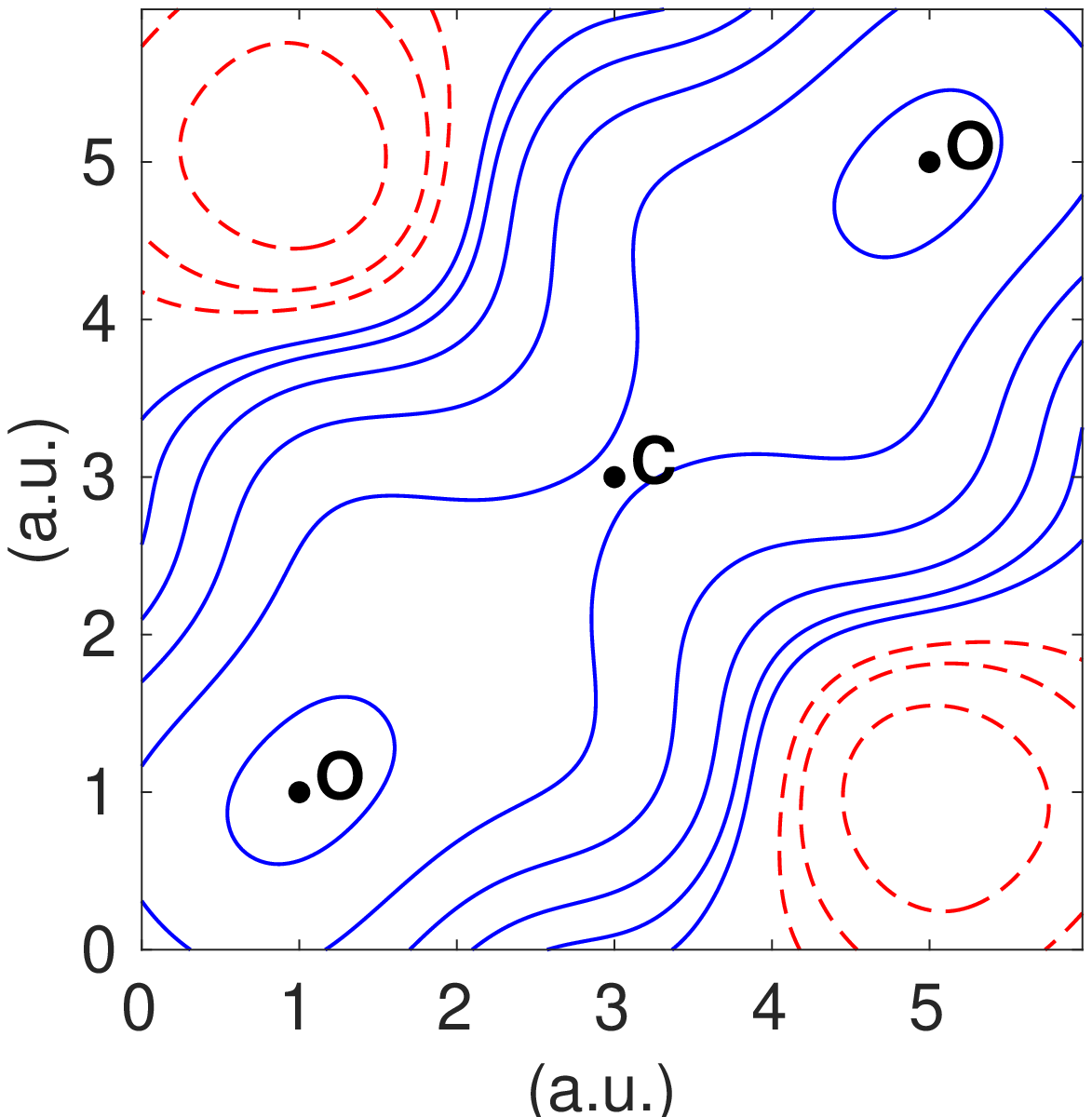}
        \caption{}
        \label{fig:RDMCrystalUp}
    \end{subfigure}
 \caption{Spin-traced 1-RDM $\widehat{\Gamma}(\b{r},\b{r'})$ contour maps for two different segments \cite{footnoteRDM}. For each segment, the position vectors $\b{r}$ (horizontal axis) and $\b{r'}$ (vertical axis) are restricted to vary along the segment. Upper panel: along the O-C-O bonding. Lower panel: along a segment parallel and $1$ a.u. away from the O-C-O bonding. Left column: inferred from position and momentum space expectation values. Right column: periodic \textit{ab-initio} computation. Contours at intervals of $\pm 0.01\times 2^n$ a.u.$^{-3}$ ($n=0$-$20$): positive and negative contours are blue solid lines and red dashed lines respectively.}
\label{fig:RDM}
\end{figure*}

\subsection{Independent molecule model}

As the four CO$_2$ molecules in the unit cell are identical and sufficiently distant from each other, each molecule can be described by the same molecular spin-traced 1-RDM in a different orientation set of local axes. Consequently, the total structure factors $ F_{tot}(\b{q})$ and directional Compton profiles $ J^{\b{u}}_{tot}(q)$ can be computed from the molecular structure factors and directional Compton profiles $F(\b{q})$ and $J^{\b{u}}(q)$ by:
\begin{align}
  F_{tot}(\b{q}) =& F(\b{q}) + \sum_{m=2}^4   e^{-i\, (\widehat{\Omega}_m\b{q})\cdot{\b{r_m}}} F(\widehat{\Omega}_m\b{q})\\
   J^{\b{u}}_{tot}(q) =&  J^{\b{u}}(q) +   \sum_{m=2}^4   J^{(\widehat{\Omega}_m\b{u})}(q)
\end{align}
where $\b{r_m}$ and $\widehat{\Omega}_m$ are respectively, the translation vector and the inverse of the rotation matrix, bringing the first molecule to molecule $m$ ($m \in {2,3,4}$).

To assess the robustness of the method, the basis set $\{\chi_i\}_{i\in\{1,\ldots,n\}}$ used to represent the spin-traced 1-RDM has been chosen to have fewer degrees of freedom and diffuseness than the one used to generate the expectation values ($3$-$21$G(d)) \cite{binkley80,schuchardt07,feller96}.

  \begin{figure*}
    \centering
    \begin{subfigure}[t]{0.4\textwidth}
        \includegraphics[width=\textwidth]{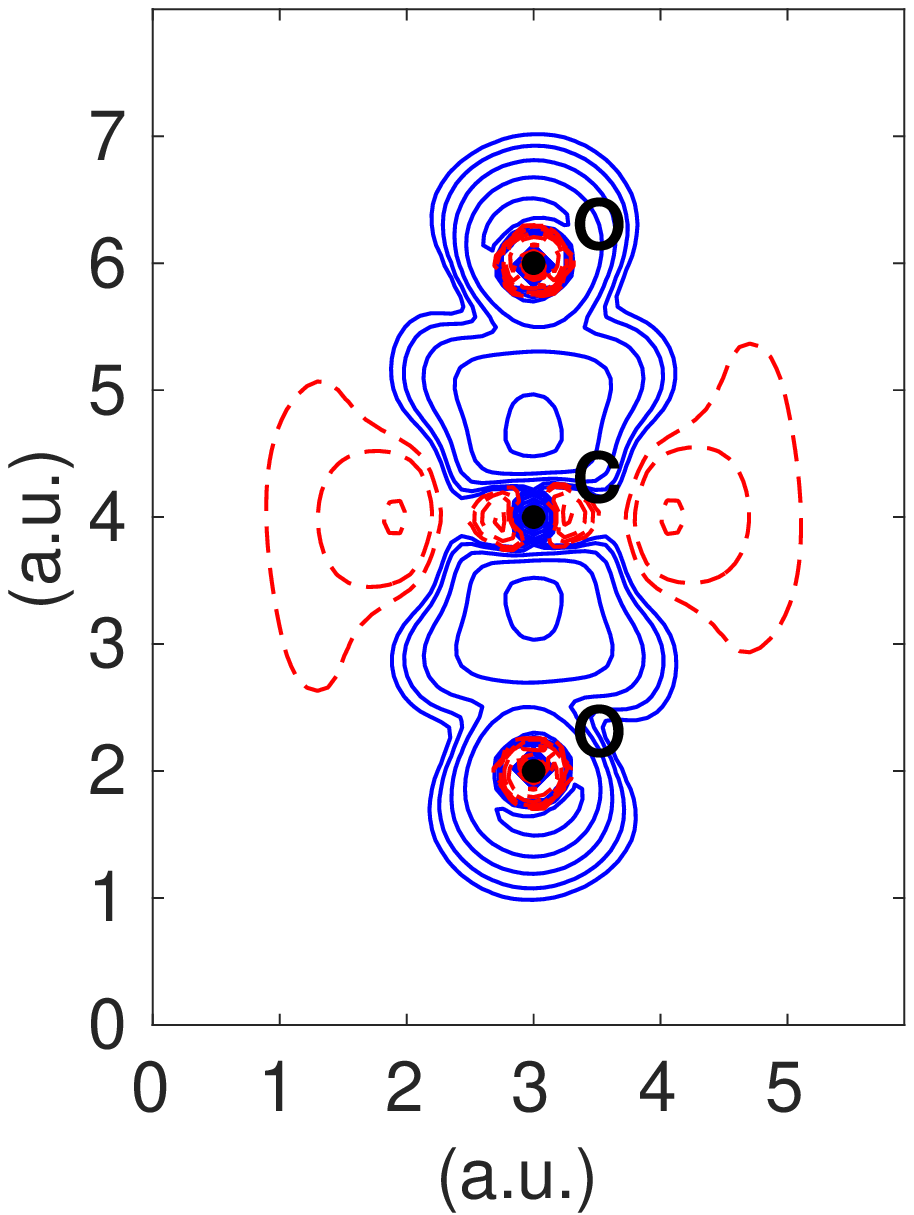}
        \caption{Inferred from position and momentum expectation values.}
        \label{fig:DENSSDP}
    \end{subfigure}
    \begin{subfigure}[t]{0.4\textwidth}
        \includegraphics[width=\textwidth]{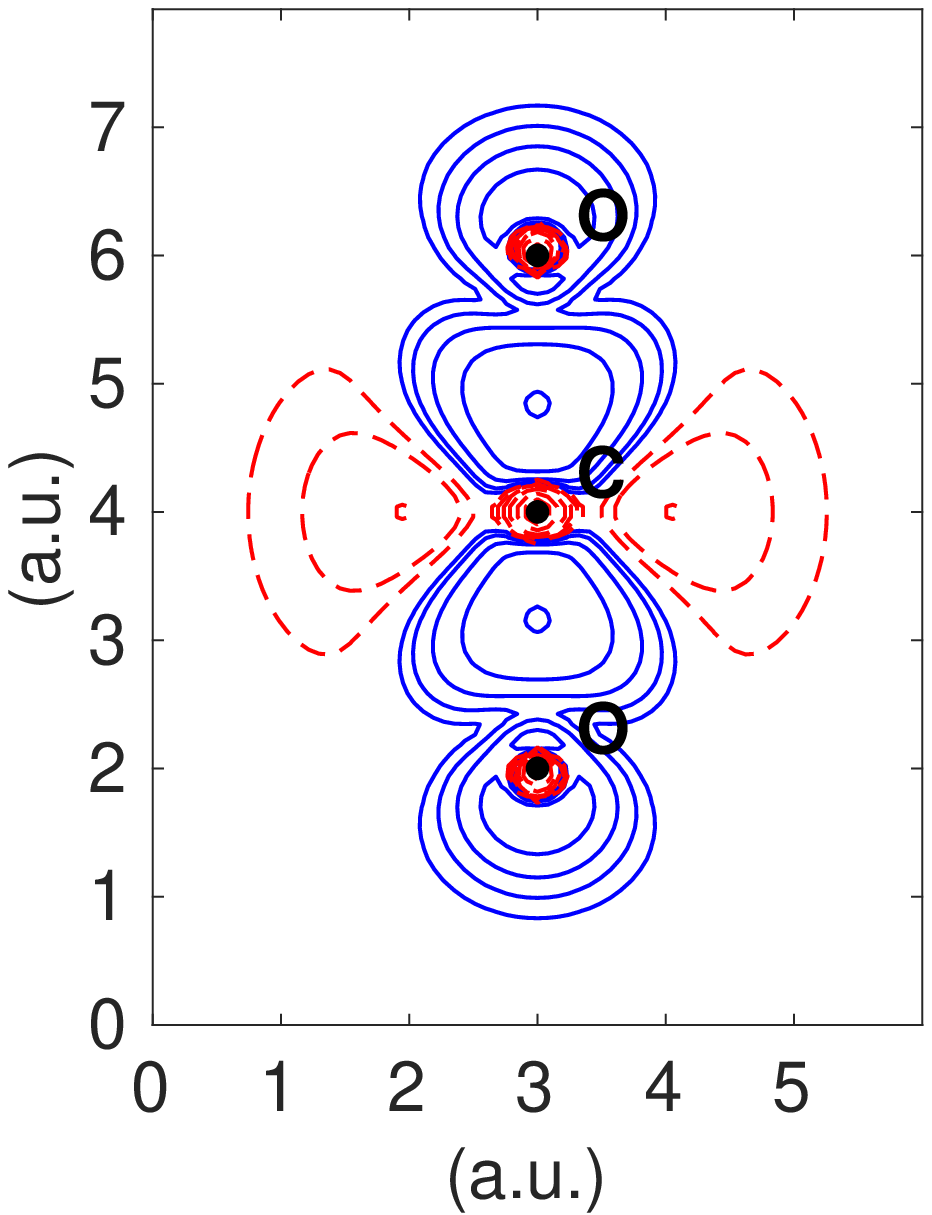}
        \caption{ Periodic \textit{ab-initio} computation.}
        \label{fig:DENSCrystal}
    \end{subfigure}
    \caption{Deformation density contour map in a plane including the O-C-O bonding. Contours at intervals of $\pm 0.01\times 2^n$ a.u.$^{-3}$ ($n=0$-$20$): positive and negative contours are blue solid lines and red dashed lines respectively.}\label{fig:DENS}
  \end{figure*}

\subsection{Results analysis}
Program (\ref{eq:s}) has been successfully solved for the case of dry ice. The DCPs and SFs computed with the optimized population matrix are near identical to their reference. In Fig.\ref{fig:CPDiff}, one DCP derived from the 1-RDM model is plotted together with its pseudo-experimental reference for comparison (see Ancillary Material for the other two DCPs). The same comparison is made for the SFs in a Fourier density map in Fig.\ref{fig:Fourier}.

The inferred and the periodic \textit{ab-initio} spin-traced 1-RDM are in close agreement along the O-C-O bond (Fig.\ref{fig:RDM}). Although slight differences are observed in the off-diagonal regions, corresponding to the subtle interactions between both bonds, the general features have been accurately reproduced.

In a plane comprising of the atoms of the molecule, the overall expected picture of the deformation density map i.e.\ the difference between the total density and the non-interacting atom density, is recovered with minor discrepancies on the oxygen atoms and around the carbon atom  (Fig.\ref{fig:DENS}). The fact that the axial symmetry is not obtained originates from the lack of symmetry constraints and the limited amount of experimental information (Fig.\ref{fig:Fourier}). It could possibly be recovered by providing additional knowledge (symmetry constraints) to the model or using more expectation values.

The off-diagonal regions in Fig.\ref{fig:RDM} are highly sensitive to the amount of noise added to the DCPs and the sharp contrast around the O-O interaction (region $5$ a.u. - $1$ a.u.) is quickly lost as the standard deviation is increased. This sensitivity might be particularly high for the case of dry ice as limited information can be deduced from DCPs because of their relatively low anistropies. Additionally, as the noise added to the SFs grow, further discrepancies appear quite naturally on the deformation density map.

\begin{figure*}
    \centering
    \begin{subfigure}[t]{0.40\textwidth}
        \includegraphics[width=\textwidth]{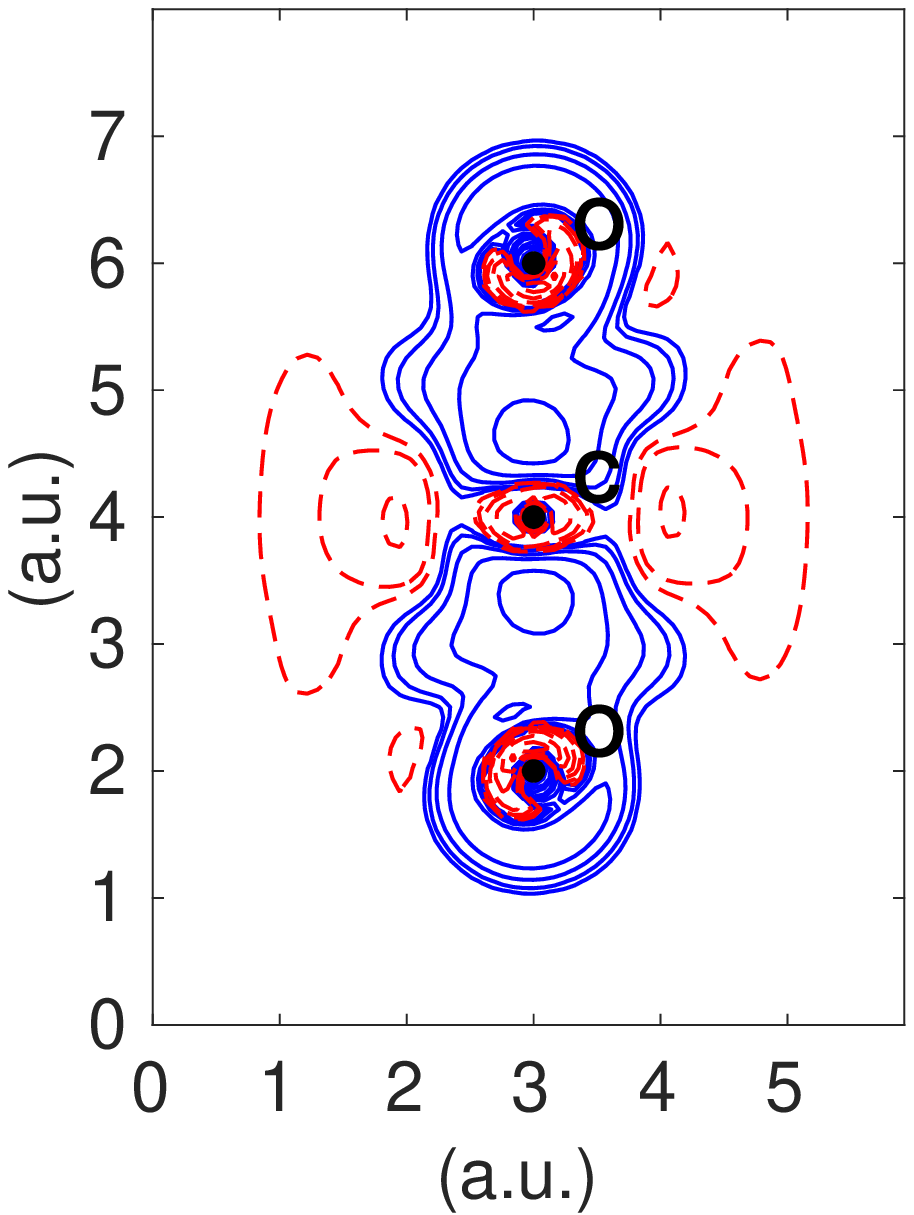}
    \end{subfigure}
    \begin{subfigure}[t]{0.40\textwidth}
        \includegraphics[width=\textwidth]{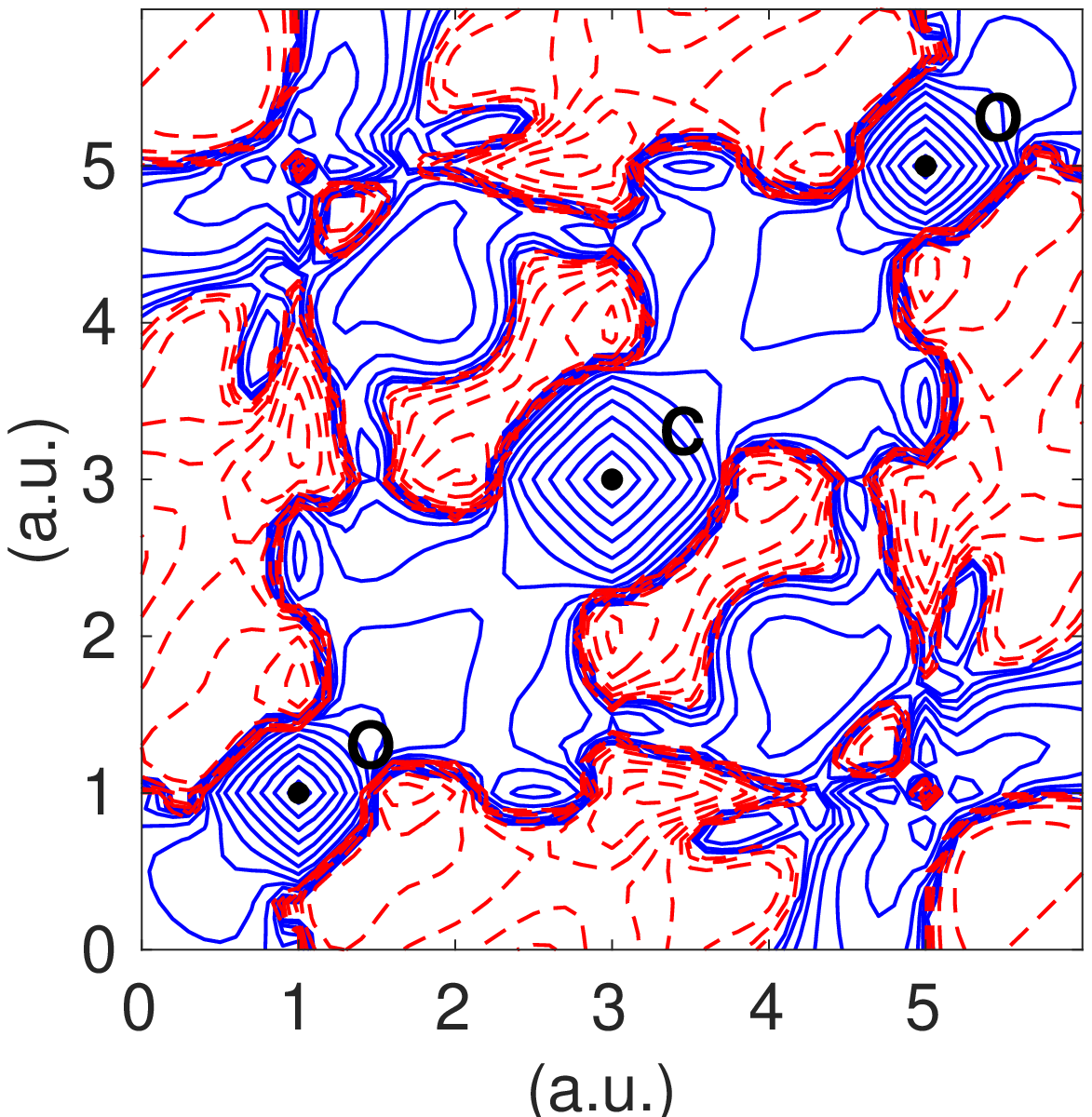}
      \end{subfigure}
    \caption{SF-only inferred deformation density map in a plane including the O-C-O bonding and spin-traced 1-RDM $\widehat{\Gamma}(\b{r},\b{r'})$ along the O-C-O bonding. Contours at intervals of $\pm 0.01\times 2^n$ a.u.$^{-3}$ ($n=0$-$20$): positive and negative contours  are blue solid lines and red dashed lines respectively.}\label{fig:CPFailure}
  \end{figure*}

Furthermore, restricting the optimization on the SFs only severely impacts the results (Fig.\ref{fig:CPFailure}) and therefore clearly illustrates the complementarity of both momentum and position expectation values as mentioned in Sec.\ref{sec:expectVal}. Of course, restricting the optimization on the DCPs gives an even worse result.

\section{Conclusion}
\label{sec:conclusion}
With the aim of inferring a 1-RDM from structure factors and directional Compton Profiles with minimal \textit{prior} knowledge, a method based on Semidefinite Programming was proposed. The effectiveness of this method has been evaluated on the crystal of dry ice taking periodic \textit{ab-initio} calculations as reference. In this example, the method was in very good agreement with the reference, showing that the use of both structure factors and directional Compton profiles provides sufficient information to infer the 1-RDM in a given atomic basis set.

Such a method could be used as a reference to interpret experimental results. For now, it is only applicable to molecular crystals but it could possibly, in the future, be extended to the modeling of 1-RDM of more general crystalline systems.

While this method is quite general, it still depends on the choice of the atomic basis set. Its result can be refined through optimization of the basis, such as in \cite{gueddida18}, however the most ideal solution would be an inference process that does not require a basis set altogether. At this stage, further work is required to achieve this as, to the best of our knowledge, the \textit{N-representability} conditions are more conveniently expressed in a given basis.

\section{Acknowledgments}
 The authors gratefully acknowledge Z. Yan whose program was used to compute the periodic \textit{ab-initio} 1-RDM along segments. Special thanks are also addressed to S. Gueddida, P. Becker, N. Ghermani for invaluable help and suggestions. B.D. wishes to acknowledge G. Valmorbida for his time explaining Semidefinite Programming and  X. Adriaens, J-Y Raty for practical suggestions and comments. J-M.G. thanks B. Gillon, M. Souhassou, N. Claiser and C. Lecomte for fruitful discussions and experimental issues. We express our warm thanks to Julie McDonald and Michelle Seeto for carefully proofreading the English in this paper. The computing cluster of CentraleSupélec has been used for this work.

\bibliography{biblio}

\providecommand{\noopsort}[1]{}\providecommand{\singleletter}[1]{#1}%
\begin{thebibliography}{42}%
\makeatletter
\providecommand \@ifxundefined [1]{%
 \@ifx{#1\undefined}
}%
\providecommand \@ifnum [1]{%
 \ifnum #1\expandafter \@firstoftwo
 \else \expandafter \@secondoftwo
 \fi
}%
\providecommand \@ifx [1]{%
 \ifx #1\expandafter \@firstoftwo
 \else \expandafter \@secondoftwo
 \fi
}%
\providecommand \natexlab [1]{#1}%
\providecommand \enquote  [1]{``#1''}%
\providecommand \bibnamefont  [1]{#1}%
\providecommand \bibfnamefont [1]{#1}%
\providecommand \citenamefont [1]{#1}%
\providecommand \href@noop [0]{\@secondoftwo}%
\providecommand \href [0]{\begingroup \@sanitize@url \@href}%
\providecommand \@href[1]{\@@startlink{#1}\@@href}%
\providecommand \@@href[1]{\endgroup#1\@@endlink}%
\providecommand \@sanitize@url [0]{\catcode `\\12\catcode `\$12\catcode
  `\&12\catcode `\#12\catcode `\^12\catcode `\_12\catcode `\%12\relax}%
\providecommand \@@startlink[1]{}%
\providecommand \@@endlink[0]{}%
\providecommand \url  [0]{\begingroup\@sanitize@url \@url }%
\providecommand \@url [1]{\endgroup\@href {#1}{\urlprefix }}%
\providecommand \urlprefix  [0]{URL }%
\providecommand \Eprint [0]{\href }%
\providecommand \doibase [0]{http://dx.doi.org/}%
\providecommand \selectlanguage [0]{\@gobble}%
\providecommand \bibinfo  [0]{\@secondoftwo}%
\providecommand \bibfield  [0]{\@secondoftwo}%
\providecommand \translation [1]{[#1]}%
\providecommand \BibitemOpen [0]{}%
\providecommand \bibitemStop [0]{}%
\providecommand \bibitemNoStop [0]{.\EOS\space}%
\providecommand \EOS [0]{\spacefactor3000\relax}%
\providecommand \BibitemShut  [1]{\csname bibitem#1\endcsname}%
\let\auto@bib@innerbib\@empty
\bibitem [{\citenamefont {L\"owdin}(1955)}]{lowdin55}%
  \BibitemOpen
  \bibfield  {author} {\bibinfo {author} {\bibfnamefont {P.-O.}\ \bibnamefont
  {L\"owdin}},\ }\href {\doibase 10.1103/PhysRev.97.1474} {\bibfield  {journal}
  {\bibinfo  {journal} {Phys. Rev.}\ }\textbf {\bibinfo {volume} {97}},\
  \bibinfo {pages} {1474} (\bibinfo {year} {1955})}\BibitemShut {NoStop}%
\bibitem [{\citenamefont {Coleman}(1963)}]{coleman63}%
  \BibitemOpen
  \bibfield  {author} {\bibinfo {author} {\bibfnamefont {A.~J.}\ \bibnamefont
  {Coleman}},\ }\href {\doibase 10.1103/RevModPhys.35.668} {\bibfield
  {journal} {\bibinfo  {journal} {Rev. Mod. Phys.}\ }\textbf {\bibinfo {volume}
  {35}},\ \bibinfo {pages} {668} (\bibinfo {year} {1963})}\BibitemShut
  {NoStop}%
\bibitem [{\citenamefont {McWeeny}(1960)}]{mcweeny60}%
  \BibitemOpen
  \bibfield  {author} {\bibinfo {author} {\bibfnamefont {R.}~\bibnamefont
  {McWeeny}},\ }\href {\doibase 10.1103/RevModPhys.32.335} {\bibfield
  {journal} {\bibinfo  {journal} {Rev. Mod. Phys.}\ }\textbf {\bibinfo {volume}
  {32}},\ \bibinfo {pages} {335} (\bibinfo {year} {1960})}\BibitemShut
  {NoStop}%
\bibitem [{\citenamefont {Davidson}(1976)}]{davidson76}%
  \BibitemOpen
  \bibfield  {author} {\bibinfo {author} {\bibfnamefont {E.}~\bibnamefont
  {Davidson}},\ }\href@noop {} {\bibfield  {journal} {\bibinfo  {journal}
  {Quantum Chemistry, Academic Press, New York}\ } (\bibinfo {year}
  {1976})}\BibitemShut {NoStop}%
\bibitem [{\citenamefont {Lathiotakis}\ and\ \citenamefont
  {Marques}(2008)}]{lathiotakis08}%
  \BibitemOpen
  \bibfield  {author} {\bibinfo {author} {\bibfnamefont {N.~N.}\ \bibnamefont
  {Lathiotakis}}\ and\ \bibinfo {author} {\bibfnamefont {M.~A.~L.}\
  \bibnamefont {Marques}},\ }\href {\doibase 10.1063/1.2899328} {\bibfield
  {journal} {\bibinfo  {journal} {The Journal of Chemical Physics}\ }\textbf
  {\bibinfo {volume} {128}},\ \bibinfo {pages} {184103} (\bibinfo {year}
  {2008})}\BibitemShut {NoStop}%
\bibitem [{\citenamefont {Gilbert}(1975)}]{gilbert75}%
  \BibitemOpen
  \bibfield  {author} {\bibinfo {author} {\bibfnamefont {T.~L.}\ \bibnamefont
  {Gilbert}},\ }\href {\doibase 10.1103/PhysRevB.12.2111} {\bibfield  {journal}
  {\bibinfo  {journal} {Phys. Rev. B}\ }\textbf {\bibinfo {volume} {12}},\
  \bibinfo {pages} {2111} (\bibinfo {year} {1975})}\BibitemShut {NoStop}%
\bibitem [{\citenamefont {Deutsch}\ \emph {et~al.}(2012)\citenamefont
  {Deutsch}, \citenamefont {Claiser}, \citenamefont {Pillet}, \citenamefont
  {Chumakov}, \citenamefont {Becker}, \citenamefont {Gillet}, \citenamefont
  {Gillon}, \citenamefont {Lecomte},\ and\ \citenamefont
  {Souhassou}}]{deutsch12}%
  \BibitemOpen
  \bibfield  {author} {\bibinfo {author} {\bibfnamefont {M.}~\bibnamefont
  {Deutsch}}, \bibinfo {author} {\bibfnamefont {N.}~\bibnamefont {Claiser}},
  \bibinfo {author} {\bibfnamefont {S.}~\bibnamefont {Pillet}}, \bibinfo
  {author} {\bibfnamefont {Y.}~\bibnamefont {Chumakov}}, \bibinfo {author}
  {\bibfnamefont {P.}~\bibnamefont {Becker}}, \bibinfo {author} {\bibfnamefont
  {J.-M.}\ \bibnamefont {Gillet}}, \bibinfo {author} {\bibfnamefont
  {B.}~\bibnamefont {Gillon}}, \bibinfo {author} {\bibfnamefont
  {C.}~\bibnamefont {Lecomte}}, \ and\ \bibinfo {author} {\bibfnamefont
  {M.}~\bibnamefont {Souhassou}},\ }\href {\doibase 10.1107/S0108767312031996}
  {\bibfield  {journal} {\bibinfo  {journal} {Acta Crystallographica Section
  A}\ }\textbf {\bibinfo {volume} {68}},\ \bibinfo {pages} {675} (\bibinfo
  {year} {2012})}\BibitemShut {NoStop}%
\bibitem [{\citenamefont {Deutsch}\ \emph {et~al.}(2014)\citenamefont
  {Deutsch}, \citenamefont {Gillon}, \citenamefont {Claiser}, \citenamefont
  {Gillet}, \citenamefont {Lecomte},\ and\ \citenamefont
  {Souhassou}}]{deutsch14}%
  \BibitemOpen
  \bibfield  {author} {\bibinfo {author} {\bibfnamefont {M.}~\bibnamefont
  {Deutsch}}, \bibinfo {author} {\bibfnamefont {B.}~\bibnamefont {Gillon}},
  \bibinfo {author} {\bibfnamefont {N.}~\bibnamefont {Claiser}}, \bibinfo
  {author} {\bibfnamefont {J.-M.}\ \bibnamefont {Gillet}}, \bibinfo {author}
  {\bibfnamefont {C.}~\bibnamefont {Lecomte}}, \ and\ \bibinfo {author}
  {\bibfnamefont {M.}~\bibnamefont {Souhassou}},\ }\href {\doibase
  10.1107/S2052252514007283} {\bibfield  {journal} {\bibinfo  {journal}
  {IUCrJ}\ }\textbf {\bibinfo {volume} {1}},\ \bibinfo {pages} {194} (\bibinfo
  {year} {2014})}\BibitemShut {NoStop}%
\bibitem [{\citenamefont {Hansen}\ and\ \citenamefont
  {Coppens}(1978)}]{hansen78}%
  \BibitemOpen
  \bibfield  {author} {\bibinfo {author} {\bibfnamefont {N.~K.}\ \bibnamefont
  {Hansen}}\ and\ \bibinfo {author} {\bibfnamefont {P.}~\bibnamefont
  {Coppens}},\ }\href {\doibase 10.1107/S0567739478001886} {\bibfield
  {journal} {\bibinfo  {journal} {Acta Crystallographica Section A}\ }\textbf
  {\bibinfo {volume} {34}},\ \bibinfo {pages} {909} (\bibinfo {year}
  {1978})}\BibitemShut {NoStop}%
\bibitem [{\citenamefont {Gillet}\ \emph {et~al.}(2001)\citenamefont {Gillet},
  \citenamefont {Becker},\ and\ \citenamefont {Cortona}}]{gillet01}%
  \BibitemOpen
  \bibfield  {author} {\bibinfo {author} {\bibfnamefont {J.-M.}\ \bibnamefont
  {Gillet}}, \bibinfo {author} {\bibfnamefont {P.~J.}\ \bibnamefont {Becker}},
  \ and\ \bibinfo {author} {\bibfnamefont {P.}~\bibnamefont {Cortona}},\ }\href
  {\doibase 10.1103/PhysRevB.63.235115} {\bibfield  {journal} {\bibinfo
  {journal} {Phys. Rev. B}\ }\textbf {\bibinfo {volume} {63}},\ \bibinfo
  {pages} {235115} (\bibinfo {year} {2001})}\BibitemShut {NoStop}%
\bibitem [{\citenamefont {Gillet}(2007)}]{gillet07}%
  \BibitemOpen
  \bibfield  {author} {\bibinfo {author} {\bibfnamefont {J.-M.}\ \bibnamefont
  {Gillet}},\ }\href {\doibase 10.1107/S0108767307001663} {\bibfield  {journal}
  {\bibinfo  {journal} {Acta Crystallographica Section A}\ }\textbf {\bibinfo
  {volume} {63}},\ \bibinfo {pages} {234} (\bibinfo {year} {2007})}\BibitemShut
  {NoStop}%
\bibitem [{\citenamefont {Gillet}\ and\ \citenamefont
  {Becker}(2004)}]{gillet04}%
  \BibitemOpen
  \bibfield  {author} {\bibinfo {author} {\bibfnamefont {J.-M.}\ \bibnamefont
  {Gillet}}\ and\ \bibinfo {author} {\bibfnamefont {P.~J.}\ \bibnamefont
  {Becker}},\ }\href {\doibase https://doi.org/10.1016/j.jpcs.2004.08.014}
  {\bibfield  {journal} {\bibinfo  {journal} {Journal of Physics and Chemistry
  of Solids}\ }\textbf {\bibinfo {volume} {65}},\ \bibinfo {pages} {2017 }
  (\bibinfo {year} {2004})},\ \bibinfo {note} {sagamore XIV: Charge, Spin and
  Momentum Densities}\BibitemShut {NoStop}%
\bibitem [{\citenamefont {Gillet}\ \emph {et~al.}(1999)\citenamefont {Gillet},
  \citenamefont {Fluteaux},\ and\ \citenamefont {Becker}}]{gillet99}%
  \BibitemOpen
  \bibfield  {author} {\bibinfo {author} {\bibfnamefont {J.-M.}\ \bibnamefont
  {Gillet}}, \bibinfo {author} {\bibfnamefont {C.}~\bibnamefont {Fluteaux}}, \
  and\ \bibinfo {author} {\bibfnamefont {P.~J.}\ \bibnamefont {Becker}},\
  }\href {\doibase 10.1103/PhysRevB.60.2345} {\bibfield  {journal} {\bibinfo
  {journal} {Phys. Rev. B}\ }\textbf {\bibinfo {volume} {60}},\ \bibinfo
  {pages} {2345} (\bibinfo {year} {1999})}\BibitemShut {NoStop}%
\bibitem [{\citenamefont {Gueddida}\ \emph {et~al.}(2018)\citenamefont
  {Gueddida}, \citenamefont {Yan}, \citenamefont {Kibalin}, \citenamefont
  {Voufack}, \citenamefont {Claiser}, \citenamefont {Souhassou}, \citenamefont
  {Lecomte}, \citenamefont {Gillon},\ and\ \citenamefont
  {Gillet}}]{gueddida18}%
  \BibitemOpen
  \bibfield  {author} {\bibinfo {author} {\bibfnamefont {S.}~\bibnamefont
  {Gueddida}}, \bibinfo {author} {\bibfnamefont {Z.}~\bibnamefont {Yan}},
  \bibinfo {author} {\bibfnamefont {I.}~\bibnamefont {Kibalin}}, \bibinfo
  {author} {\bibfnamefont {A.~B.}\ \bibnamefont {Voufack}}, \bibinfo {author}
  {\bibfnamefont {N.}~\bibnamefont {Claiser}}, \bibinfo {author} {\bibfnamefont
  {M.}~\bibnamefont {Souhassou}}, \bibinfo {author} {\bibfnamefont
  {C.}~\bibnamefont {Lecomte}}, \bibinfo {author} {\bibfnamefont
  {B.}~\bibnamefont {Gillon}}, \ and\ \bibinfo {author} {\bibfnamefont {J.-M.}\
  \bibnamefont {Gillet}},\ }\href@noop {} {\bibfield  {journal} {\bibinfo
  {journal} {The Journal of chemical physics}\ }\textbf {\bibinfo {volume}
  {148}},\ \bibinfo {pages} {164106} (\bibinfo {year} {2018})}\BibitemShut
  {NoStop}%
\bibitem [{\citenamefont {Pillet}\ \emph {et~al.}(2001)\citenamefont {Pillet},
  \citenamefont {Souhassou}, \citenamefont {Pontillon}, \citenamefont
  {Caneschi}, \citenamefont {Gatteschi},\ and\ \citenamefont
  {Lecomte}}]{pillet01}%
  \BibitemOpen
  \bibfield  {author} {\bibinfo {author} {\bibfnamefont {S.}~\bibnamefont
  {Pillet}}, \bibinfo {author} {\bibfnamefont {M.}~\bibnamefont {Souhassou}},
  \bibinfo {author} {\bibfnamefont {Y.}~\bibnamefont {Pontillon}}, \bibinfo
  {author} {\bibfnamefont {A.}~\bibnamefont {Caneschi}}, \bibinfo {author}
  {\bibfnamefont {D.}~\bibnamefont {Gatteschi}}, \ and\ \bibinfo {author}
  {\bibfnamefont {C.}~\bibnamefont {Lecomte}},\ }\href {\doibase
  10.1039/B003674I} {\bibfield  {journal} {\bibinfo  {journal} {New J. Chem.}\
  }\textbf {\bibinfo {volume} {25}},\ \bibinfo {pages} {131} (\bibinfo {year}
  {2001})}\BibitemShut {NoStop}%
\bibitem [{\citenamefont {Schmider}\ \emph {et~al.}(1992)\citenamefont
  {Schmider}, \citenamefont {Smith},\ and\ \citenamefont
  {Weyrich}}]{schmider92}%
  \BibitemOpen
  \bibfield  {author} {\bibinfo {author} {\bibfnamefont {H.}~\bibnamefont
  {Schmider}}, \bibinfo {author} {\bibfnamefont {V.~H.}\ \bibnamefont {Smith}},
  \ and\ \bibinfo {author} {\bibfnamefont {W.}~\bibnamefont {Weyrich}},\ }\href
  {\doibase 10.1063/1.462256} {\bibfield  {journal} {\bibinfo  {journal} {The
  Journal of Chemical Physics}\ }\textbf {\bibinfo {volume} {96}},\ \bibinfo
  {pages} {8986} (\bibinfo {year} {1992})}\BibitemShut {NoStop}%
\bibitem [{\citenamefont {Clinton}\ and\ \citenamefont
  {Massa}(1972)}]{clinton72}%
  \BibitemOpen
  \bibfield  {author} {\bibinfo {author} {\bibfnamefont {W.~L.}\ \bibnamefont
  {Clinton}}\ and\ \bibinfo {author} {\bibfnamefont {L.~J.}\ \bibnamefont
  {Massa}},\ }\href {\doibase 10.1103/PhysRevLett.29.1363} {\bibfield
  {journal} {\bibinfo  {journal} {Phys. Rev. Lett.}\ }\textbf {\bibinfo
  {volume} {29}},\ \bibinfo {pages} {1363} (\bibinfo {year}
  {1972})}\BibitemShut {NoStop}%
\bibitem [{\citenamefont {Tsirelson}\ and\ \citenamefont
  {Ozerov}(1996)}]{tsirelson96}%
  \BibitemOpen
  \bibfield  {author} {\bibinfo {author} {\bibfnamefont {V.~G.}\ \bibnamefont
  {Tsirelson}}\ and\ \bibinfo {author} {\bibfnamefont {R.~P.}\ \bibnamefont
  {Ozerov}},\ }\href@noop {} {\emph {\bibinfo {title} {Electron Density and
  Bonding in Crystals: Principles, Theory and X-ray Diffraction Experiments in
  Solid State Physics and Chemistry}}}\ (\bibinfo  {publisher} {CRC Press},\
  \bibinfo {year} {1996})\BibitemShut {NoStop}%
\bibitem [{\citenamefont {Cooper}\ \emph {et~al.}(2004)\citenamefont {Cooper},
  \citenamefont {Cooper}, \citenamefont {Mijnarends}, \citenamefont
  {Mijnarends}, \citenamefont {Shiotani}, \citenamefont {Sakai},\ and\
  \citenamefont {Bansil}}]{cooper04}%
  \BibitemOpen
  \bibfield  {author} {\bibinfo {author} {\bibfnamefont {M.~J.}\ \bibnamefont
  {Cooper}}, \bibinfo {author} {\bibfnamefont {M.}~\bibnamefont {Cooper}},
  \bibinfo {author} {\bibfnamefont {P.~E.}\ \bibnamefont {Mijnarends}},
  \bibinfo {author} {\bibfnamefont {P.}~\bibnamefont {Mijnarends}}, \bibinfo
  {author} {\bibfnamefont {N.}~\bibnamefont {Shiotani}}, \bibinfo {author}
  {\bibfnamefont {N.}~\bibnamefont {Sakai}}, \ and\ \bibinfo {author}
  {\bibfnamefont {A.}~\bibnamefont {Bansil}},\ }\href@noop {} {\emph {\bibinfo
  {title} {X-ray Compton scattering}}},\ \bibinfo {number} {5}\ (\bibinfo
  {publisher} {Oxford University Press on Demand},\ \bibinfo {year}
  {2004})\BibitemShut {NoStop}%
\bibitem [{\citenamefont {Pisani}(2012)}]{pisani12}%
  \BibitemOpen
  \bibfield  {author} {\bibinfo {author} {\bibfnamefont {C.}~\bibnamefont
  {Pisani}},\ }\href@noop {} {\emph {\bibinfo {title} {Quantum-mechanical
  ab-initio calculation of the properties of crystalline materials}}},\
  Vol.~\bibinfo {volume} {67}\ (\bibinfo  {publisher} {Springer Science \&
  Business Media},\ \bibinfo {year} {2012})\BibitemShut {NoStop}%
\bibitem [{\citenamefont {Vandenberghe}\ and\ \citenamefont
  {Boyd}(1996)}]{boyd96}%
  \BibitemOpen
  \bibfield  {author} {\bibinfo {author} {\bibfnamefont {L.}~\bibnamefont
  {Vandenberghe}}\ and\ \bibinfo {author} {\bibfnamefont {S.}~\bibnamefont
  {Boyd}},\ }\href {\doibase 10.1137/1038003} {\bibfield  {journal} {\bibinfo
  {journal} {SIAM Review}\ }\textbf {\bibinfo {volume} {38}},\ \bibinfo {pages}
  {49} (\bibinfo {year} {1996})}\BibitemShut {NoStop}%
\bibitem [{\citenamefont {Boyd}\ and\ \citenamefont
  {Vandenberghe}(2004)}]{boyd04}%
  \BibitemOpen
  \bibfield  {author} {\bibinfo {author} {\bibfnamefont {S.}~\bibnamefont
  {Boyd}}\ and\ \bibinfo {author} {\bibfnamefont {L.}~\bibnamefont
  {Vandenberghe}},\ }\href@noop {} {\emph {\bibinfo {title} {Convex
  Optimization}}}\ (\bibinfo  {publisher} {Cambridge University Press},\
  \bibinfo {address} {New York, NY, USA},\ \bibinfo {year} {2004})\BibitemShut
  {NoStop}%
\bibitem [{\citenamefont {Wolkowicz}\ \emph {et~al.}(2012)\citenamefont
  {Wolkowicz}, \citenamefont {Saigal},\ and\ \citenamefont
  {Vandenberghe}}]{wolkowicz12}%
  \BibitemOpen
  \bibfield  {author} {\bibinfo {author} {\bibfnamefont {H.}~\bibnamefont
  {Wolkowicz}}, \bibinfo {author} {\bibfnamefont {R.}~\bibnamefont {Saigal}}, \
  and\ \bibinfo {author} {\bibfnamefont {L.}~\bibnamefont {Vandenberghe}},\
  }\href@noop {} {\emph {\bibinfo {title} {Handbook of semidefinite
  programming: theory, algorithms, and applications}}},\ Vol.~\bibinfo {volume}
  {27}\ (\bibinfo  {publisher} {Springer Science \& Business Media},\ \bibinfo
  {year} {2012})\BibitemShut {NoStop}%
\bibitem [{\citenamefont {Löwdin}(1950)}]{lowdin50}%
  \BibitemOpen
  \bibfield  {author} {\bibinfo {author} {\bibfnamefont {P.}~\bibnamefont
  {Löwdin}},\ }\href {\doibase 10.1063/1.1747632} {\bibfield  {journal}
  {\bibinfo  {journal} {The Journal of Chemical Physics}\ }\textbf {\bibinfo
  {volume} {18}},\ \bibinfo {pages} {365} (\bibinfo {year} {1950})}\BibitemShut
  {NoStop}%
\bibitem [{\citenamefont {Pattison}\ and\ \citenamefont
  {Weyrich}(1979)}]{pattison79}%
  \BibitemOpen
  \bibfield  {author} {\bibinfo {author} {\bibfnamefont {P.}~\bibnamefont
  {Pattison}}\ and\ \bibinfo {author} {\bibfnamefont {W.}~\bibnamefont
  {Weyrich}},\ }\href {\doibase https://doi.org/10.1016/0022-3697(79)90018-0}
  {\bibfield  {journal} {\bibinfo  {journal} {Journal of Physics and Chemistry
  of Solids}\ }\textbf {\bibinfo {volume} {40}},\ \bibinfo {pages} {213 }
  (\bibinfo {year} {1979})}\BibitemShut {NoStop}%
\bibitem [{\citenamefont {Weyrich}\ \emph {et~al.}(1979)\citenamefont
  {Weyrich}, \citenamefont {Pattison},\ and\ \citenamefont
  {Williams}}]{weyrich79}%
  \BibitemOpen
  \bibfield  {author} {\bibinfo {author} {\bibfnamefont {W.}~\bibnamefont
  {Weyrich}}, \bibinfo {author} {\bibfnamefont {P.}~\bibnamefont {Pattison}}, \
  and\ \bibinfo {author} {\bibfnamefont {B.}~\bibnamefont {Williams}},\ }\href
  {\doibase https://doi.org/10.1016/0301-0104(79)80034-8} {\bibfield  {journal}
  {\bibinfo  {journal} {Chemical Physics}\ }\textbf {\bibinfo {volume} {41}},\
  \bibinfo {pages} {271 } (\bibinfo {year} {1979})}\BibitemShut {NoStop}%
\bibitem [{\citenamefont {Benesch}\ \emph {et~al.}(1971)\citenamefont
  {Benesch}, \citenamefont {Singh},\ and\ \citenamefont
  {Smith~Jr}}]{benesch71}%
  \BibitemOpen
  \bibfield  {author} {\bibinfo {author} {\bibfnamefont {R.}~\bibnamefont
  {Benesch}}, \bibinfo {author} {\bibfnamefont {S.}~\bibnamefont {Singh}}, \
  and\ \bibinfo {author} {\bibfnamefont {V.}~\bibnamefont {Smith~Jr}},\
  }\href@noop {} {\bibfield  {journal} {\bibinfo  {journal} {Chemical Physics
  Letters}\ }\textbf {\bibinfo {volume} {10}},\ \bibinfo {pages} {151}
  (\bibinfo {year} {1971})}\BibitemShut {NoStop}%
\bibitem [{\citenamefont {Sivia}\ and\ \citenamefont
  {Skilling}(2006)}]{sivia06}%
  \BibitemOpen
  \bibfield  {author} {\bibinfo {author} {\bibfnamefont {D.}~\bibnamefont
  {Sivia}}\ and\ \bibinfo {author} {\bibfnamefont {J.}~\bibnamefont
  {Skilling}},\ }\href@noop {} {\emph {\bibinfo {title} {Data analysis: a
  Bayesian tutorial}}}\ (\bibinfo  {publisher} {OUP Oxford},\ \bibinfo {year}
  {2006})\BibitemShut {NoStop}%
\bibitem [{\citenamefont {Zhang}(2006)}]{zhang06}%
  \BibitemOpen
  \bibfield  {author} {\bibinfo {author} {\bibfnamefont {F.}~\bibnamefont
  {Zhang}},\ }\href@noop {} {\emph {\bibinfo {title} {The Schur complement and
  its applications}}},\ Vol.~\bibinfo {volume} {4}\ (\bibinfo  {publisher}
  {Springer Science \& Business Media},\ \bibinfo {year} {2006})\BibitemShut
  {NoStop}%
\bibitem [{\citenamefont {Mazziotti}(2007)}]{mazziotti07}%
  \BibitemOpen
  \bibfield  {author} {\bibinfo {author} {\bibfnamefont {D.~A.}\ \bibnamefont
  {Mazziotti}},\ }\href@noop {} {\bibfield  {journal} {\bibinfo  {journal}
  {ESAIM: Mathematical Modelling and Numerical Analysis}\ }\textbf {\bibinfo
  {volume} {41}},\ \bibinfo {pages} {249} (\bibinfo {year} {2007})}\BibitemShut
  {NoStop}%
\bibitem [{\citenamefont {ApS}(2017)}]{mosek}%
  \BibitemOpen
  \bibfield  {author} {\bibinfo {author} {\bibfnamefont {M.}~\bibnamefont
  {ApS}},\ }\href {http://docs.mosek.com/8.1/toolbox/index.html} {\emph
  {\bibinfo {title} {The MOSEK optimization toolbox for MATLAB manual. Version
  8.1.}}} (\bibinfo {year} {2017})\BibitemShut {NoStop}%
\bibitem [{\citenamefont {L{\"{o}}fberg}(2004)}]{Lofberg2004}%
  \BibitemOpen
  \bibfield  {author} {\bibinfo {author} {\bibfnamefont {J.}~\bibnamefont
  {L{\"{o}}fberg}},\ }in\ \href@noop {} {\emph {\bibinfo {booktitle} {In
  Proceedings of the CACSD Conference}}}\ (\bibinfo {address} {Taipei,
  Taiwan},\ \bibinfo {year} {2004})\BibitemShut {NoStop}%
\bibitem [{\citenamefont {De~Smedt}\ and\ \citenamefont
  {Keesom}(1924)}]{deSmedt1924}%
  \BibitemOpen
  \bibfield  {author} {\bibinfo {author} {\bibfnamefont {J.}~\bibnamefont
  {De~Smedt}}\ and\ \bibinfo {author} {\bibfnamefont {W.}~\bibnamefont
  {Keesom}},\ }in\ \href@noop {} {\emph {\bibinfo {booktitle} {Proceedings of
  the Koninklijke Akademie Van Wetenschappen Te Amsterdam}}},\ Vol.~\bibinfo
  {volume} {27}\ (\bibinfo {year} {1924})\ pp.\ \bibinfo {pages}
  {839--846}\BibitemShut {NoStop}%
\bibitem [{\citenamefont {Dovesi}\ \emph
  {et~al.}(2014{\natexlab{a}})\citenamefont {Dovesi}, \citenamefont {Orlando},
  \citenamefont {Erba}, \citenamefont {Zicovich-Wilson}, \citenamefont
  {Civalleri}, \citenamefont {Casassa}, \citenamefont {Maschio}, \citenamefont
  {Ferrabone}, \citenamefont {Pierre}, \citenamefont {D’Arco}, \citenamefont
  {Noel}, \citenamefont {Causa}, \citenamefont {Rerat},\ and\ \citenamefont
  {Kirtman.}}]{DovesiOrlando14}%
  \BibitemOpen
  \bibfield  {author} {\bibinfo {author} {\bibfnamefont {R.}~\bibnamefont
  {Dovesi}}, \bibinfo {author} {\bibfnamefont {R.}~\bibnamefont {Orlando}},
  \bibinfo {author} {\bibfnamefont {A.}~\bibnamefont {Erba}}, \bibinfo {author}
  {\bibfnamefont {C.~M.}\ \bibnamefont {Zicovich-Wilson}}, \bibinfo {author}
  {\bibfnamefont {B.}~\bibnamefont {Civalleri}}, \bibinfo {author}
  {\bibfnamefont {S.}~\bibnamefont {Casassa}}, \bibinfo {author} {\bibfnamefont
  {L.}~\bibnamefont {Maschio}}, \bibinfo {author} {\bibfnamefont
  {M.}~\bibnamefont {Ferrabone}}, \bibinfo {author} {\bibfnamefont {M.~D.~L.}\
  \bibnamefont {Pierre}}, \bibinfo {author} {\bibfnamefont {P.}~\bibnamefont
  {D’Arco}}, \bibinfo {author} {\bibfnamefont {Y.}~\bibnamefont {Noel}},
  \bibinfo {author} {\bibfnamefont {M.}~\bibnamefont {Causa}}, \bibinfo
  {author} {\bibfnamefont {M.}~\bibnamefont {Rerat}}, \ and\ \bibinfo {author}
  {\bibfnamefont {B.}~\bibnamefont {Kirtman.}},\ }\href@noop {} {\bibfield
  {journal} {\bibinfo  {journal} {Int.J. Quantum Chem.}\ }\textbf {\bibinfo
  {volume} {114}} (\bibinfo {year} {2014}{\natexlab{a}})}\BibitemShut {NoStop}%
\bibitem [{\citenamefont {Dovesi}\ \emph
  {et~al.}(2014{\natexlab{b}})\citenamefont {Dovesi}, \citenamefont {Saunders},
  \citenamefont {Roetti}, \citenamefont {Orlando}, \citenamefont
  {Zicovich-Wilson}, \citenamefont {Pascale}, \citenamefont {Civalleri},
  \citenamefont {Doll}, \citenamefont {Harrison}, \citenamefont {Bush},
  \citenamefont {D’Arco}, \citenamefont {Llunell}, \citenamefont {Causà},\
  and\ \citenamefont {Noël}}]{DovesiSaunders14}%
  \BibitemOpen
  \bibfield  {author} {\bibinfo {author} {\bibfnamefont {R.}~\bibnamefont
  {Dovesi}}, \bibinfo {author} {\bibfnamefont {V.~R.}\ \bibnamefont
  {Saunders}}, \bibinfo {author} {\bibfnamefont {C.}~\bibnamefont {Roetti}},
  \bibinfo {author} {\bibfnamefont {R.}~\bibnamefont {Orlando}}, \bibinfo
  {author} {\bibfnamefont {C.~M.}\ \bibnamefont {Zicovich-Wilson}}, \bibinfo
  {author} {\bibfnamefont {F.}~\bibnamefont {Pascale}}, \bibinfo {author}
  {\bibfnamefont {B.}~\bibnamefont {Civalleri}}, \bibinfo {author}
  {\bibfnamefont {K.}~\bibnamefont {Doll}}, \bibinfo {author} {\bibfnamefont
  {N.~M.}\ \bibnamefont {Harrison}}, \bibinfo {author} {\bibfnamefont {I.~J.}\
  \bibnamefont {Bush}}, \bibinfo {author} {\bibfnamefont {P.}~\bibnamefont
  {D’Arco}}, \bibinfo {author} {\bibfnamefont {M.}~\bibnamefont {Llunell}},
  \bibinfo {author} {\bibfnamefont {M.}~\bibnamefont {Causà}}, \ and\ \bibinfo
  {author} {\bibfnamefont {Y.}~\bibnamefont {Noël}},\ }\href@noop {} {\enquote
  {\bibinfo {title} {Crystal14 user's manual},}\ } (\bibinfo {year}
  {2014}{\natexlab{b}})\BibitemShut {NoStop}%
\bibitem [{\citenamefont {Peintinger}\ \emph {et~al.}(2013)\citenamefont
  {Peintinger}, \citenamefont {Oliveira},\ and\ \citenamefont
  {Bredow}}]{peintinger13}%
  \BibitemOpen
  \bibfield  {author} {\bibinfo {author} {\bibfnamefont {M.~F.}\ \bibnamefont
  {Peintinger}}, \bibinfo {author} {\bibfnamefont {D.~V.}\ \bibnamefont
  {Oliveira}}, \ and\ \bibinfo {author} {\bibfnamefont {T.}~\bibnamefont
  {Bredow}},\ }\href@noop {} {\bibfield  {journal} {\bibinfo  {journal}
  {Journal of Computational Chemistry}\ }\textbf {\bibinfo {volume} {34}},\
  \bibinfo {pages} {451} (\bibinfo {year} {2013})}\BibitemShut {NoStop}%
\bibitem [{\citenamefont {Civalleri}\ \emph {et~al.}(2012)\citenamefont
  {Civalleri}, \citenamefont {Presti}, \citenamefont {Dovesi},\ and\
  \citenamefont {Savin}}]{civalleri12}%
  \BibitemOpen
  \bibfield  {author} {\bibinfo {author} {\bibfnamefont {B.}~\bibnamefont
  {Civalleri}}, \bibinfo {author} {\bibfnamefont {D.}~\bibnamefont {Presti}},
  \bibinfo {author} {\bibfnamefont {R.}~\bibnamefont {Dovesi}}, \ and\ \bibinfo
  {author} {\bibfnamefont {A.}~\bibnamefont {Savin}},\ }\href@noop {}
  {\bibfield  {journal} {\bibinfo  {journal} {Chem. Modell}\ }\textbf {\bibinfo
  {volume} {9}},\ \bibinfo {pages} {168} (\bibinfo {year} {2012})}\BibitemShut
  {NoStop}%
\bibitem [{foo({\natexlab{a}})}]{footnoteFourier}%
  \BibitemOpen
  \href@noop {} {}\bibinfo {howpublished} {Although density is a non-negative
  valued function, negative regions appear, as the Fourier series representing
  density is truncated. This is not an issue as the goal is to compare the
  structure factors} \BibitemShut {NoStop}%
\bibitem [{foo({\natexlab{b}})}]{footnoteRDM}%
  \BibitemOpen
  \href@noop {} {}\bibinfo {howpublished} {As the 1-RDM is a six-variable
  function, no convenient graphical reprentation exists apart from restricting
  the variation of the two position vectors of $\Gamma(\b{r},\b{r'})$ along a
  path} \BibitemShut {NoStop}%
\bibitem [{\citenamefont {Binkley}\ \emph {et~al.}(1980)\citenamefont
  {Binkley}, \citenamefont {Pople},\ and\ \citenamefont {Hehre}}]{binkley80}%
  \BibitemOpen
  \bibfield  {author} {\bibinfo {author} {\bibfnamefont {J.~S.}\ \bibnamefont
  {Binkley}}, \bibinfo {author} {\bibfnamefont {J.~A.}\ \bibnamefont {Pople}},
  \ and\ \bibinfo {author} {\bibfnamefont {W.~J.}\ \bibnamefont {Hehre}},\
  }\href@noop {} {\bibfield  {journal} {\bibinfo  {journal} {Journal of the
  American Chemical Society}\ }\textbf {\bibinfo {volume} {102}},\ \bibinfo
  {pages} {939} (\bibinfo {year} {1980})}\BibitemShut {NoStop}%
\bibitem [{\citenamefont {Schuchardt}\ \emph {et~al.}(2007)\citenamefont
  {Schuchardt}, \citenamefont {Didier}, \citenamefont {Elsethagen},
  \citenamefont {Sun}, \citenamefont {Gurumoorthi}, \citenamefont {Chase},
  \citenamefont {Li},\ and\ \citenamefont {Windus}}]{schuchardt07}%
  \BibitemOpen
  \bibfield  {author} {\bibinfo {author} {\bibfnamefont {K.~L.}\ \bibnamefont
  {Schuchardt}}, \bibinfo {author} {\bibfnamefont {B.~T.}\ \bibnamefont
  {Didier}}, \bibinfo {author} {\bibfnamefont {T.}~\bibnamefont {Elsethagen}},
  \bibinfo {author} {\bibfnamefont {L.}~\bibnamefont {Sun}}, \bibinfo {author}
  {\bibfnamefont {V.}~\bibnamefont {Gurumoorthi}}, \bibinfo {author}
  {\bibfnamefont {J.}~\bibnamefont {Chase}}, \bibinfo {author} {\bibfnamefont
  {J.}~\bibnamefont {Li}}, \ and\ \bibinfo {author} {\bibfnamefont {T.~L.}\
  \bibnamefont {Windus}},\ }\href@noop {} {\bibfield  {journal} {\bibinfo
  {journal} {Journal of chemical information and modeling}\ }\textbf {\bibinfo
  {volume} {47}},\ \bibinfo {pages} {1045} (\bibinfo {year}
  {2007})}\BibitemShut {NoStop}%
\bibitem [{\citenamefont {Feller}(1996)}]{feller96}%
  \BibitemOpen
  \bibfield  {author} {\bibinfo {author} {\bibfnamefont {D.}~\bibnamefont
  {Feller}},\ }\href@noop {} {\bibfield  {journal} {\bibinfo  {journal}
  {Journal of computational chemistry}\ }\textbf {\bibinfo {volume} {17}},\
  \bibinfo {pages} {1571} (\bibinfo {year} {1996})}\BibitemShut {NoStop}%
\end{thebibliography}%

\end{document}